\UseRawInputEncoding
\documentclass[preprint,12pt]{elsarticle}

 \usepackage{xcolor}
 \usepackage{color}

\usepackage{amssymb}

\begin{document}

\begin{frontmatter}


 \title{Ultrahigh  flexoelectric effect of 3D interconnected porous  polymers: modelling and verification}

\cortext[cor1]{Corresponding Author: Li-Hua Shao}
 \author[1,4]{Mingyuan Zhang\footnote{\textsuperscript{1}  Contribute equally to this work.}}
  \author[1,4]{Dongze Yan}
   \author[2,3]{Jianxiang Wang}
    \author[1]{Li-Hua Shao\corref{cor1}}
     \ead{shaolihua@buaa.edu.cn}
           \address[1]{Institute of Solid Mechanics, Beihang University, Beijing 100191, P.R.China.}
           \address[2]{State Key Laboratory for Turbulence and Complex System, Department of Mechanics and Engineering Science, College of Engineering, Peking University, Beijing 100871, P.R. China.}
           \address[3]{CAPT-HEDPS, and IFSA Collaborative Innovation Center of MoE, College of Engineering, Peking University, Beijing 100871, P.R. China.}


\begin{abstract}
Non-conductive materials like rubbers, plastics, ceramics, and even semiconductors have the property of flexoelectricity, which means that they can generate electricity when bent and twisted. However, an irregular shape or a peculiar load has been the necessary condition to realize flexoelectricity, and the weight and deformability specific ratios of flexoelectricity of solids are limited. In this work, we develop a theoretical model of flexoelectricity of three-dimensional interconnected porous materials. Compared to the solid materials, porous materials can exhibit flexoelectricity under arbitrary loading forms due to their complex microstructures, and the weight and deformability specific flexoelectric output is much higher than that of the solids. Then, we verify the model by measuring the flexoelectric response of polydimethylsiloxane (PDMS) and porous polyvinylidene fluoride (PVDF). The porous PDMS with 3D micron-scale interconnected structures exhibits two orders of magnitude higher  weight and deformability specific flexoelectric output than that of the solid truncated pyramid PDMS. The flexoelectric signal is found to be linearly proportional to the applied strain, the microstructural size and the frequency. Finally, we apply the theory to a more practical bending sensor, and demonstrate its stable functioning and accurate response. Our model can be applied to other porous materials, and the results highlight the new potential of porous micro-structured materials with a significant flexoelectric effect in the fields of mechanical sensing, actuating, energy harvesting, and biomimetics as light-weight materials.

\end{abstract}

\begin{keyword}
Flexoelectricity, Porous PDMS/PVDF,  Weight and deformability specific flexoelectric output, Bending deformation.

\end{keyword}

\end{frontmatter}


\section{Introduction}

Flexoelectricity represents the electro-mechanical coupling effect between strain gradient and electric polarization (the direct effect)~\cite{00} or electric field gradient and mechanical stress (the converse effect)~\cite{1,2,3}. The flexoelectric effect exists in all dielectrics without symmetry requirement~\cite{4,5} and it is proportional to the strain gradient. Therefore, flexoelectricity is expected to be more significant in nanomaterials compared to macroscopic materials since the strain gradient is inversely proportional to the structure size~\cite{6,7}. A significant flexoelectric effect has been observed in ferroelectrics~\cite{8,9}, polymers~\cite{10,11}, bone minerals~\cite{12,13}, biomembranes~\cite{01,02,03} and semiconductors~\cite{14,15}. Flexoelectricity has demonstrated its broad application prospects in areas of sensing~\cite{16,17}, mechanical actuating~\cite{18,19}, energy harvesting~\cite{20,04} and high-density data storage~\cite{21}.
With the extensive applications of flexible electronics and wearable devices, the demand for flexible sensing devices increases fastly~\cite{22}. Flexible polymers have shown their unique advantages in sensing applications based on the flexoelectric effect due to their softness, flexibility and eco-friendliness. {red}In soft materials, non-uniform distribution of dipoles and charges can lead to electromechanical responses~\cite{05}. Especially, flexible polymers can realize large elastic deformation, and thus large strain gradients. Hence, many researchers have investigated the flexoelectric effect of flexible polymers, such as polyurethane (PU)~\cite{23}, polyethylene terephthalate (PET)~\cite{24}, polydimethylsiloxane (PDMS)~\cite{25,26} and polyvinylidene fluoride (PVDF)~\cite{27}, along with flexible polymer/ceramic-composites~\cite{25}. Based on the inverse flexoelectric effect, Deng et al. designed an impressive flexing actuator that generates large curvature on the scale of millimeters~\cite{06}. However, the flexoelectric coefficients of flexible polymers are typically three or four orders smaller~\cite{23,28} than those of ferroelectric ceramics like BST~\cite{29} and PT~\cite{8}. Nonetheless, the low density, high flexibility and easiness to fabricate of flexible polymers enable a promising enhancement of flexoelectricity through modification~\cite{25,30} or design~\cite{31,32}.
In most of the work on the flexoelectric effect, the strain gradient is realized by one or a combination of the following methods: (1) bending of a thin beam~\cite{5,8}; (2) compressing of a truncated pyramid~\cite{11,33}; (3) twisting of a cylinder~\cite{34}; (4) atomic force microscope probe excitation of a crack~\cite{6,9}. High strain gradients and thus large flexoelectric effects can also be realized by lattice mismatch in epitaxial film structures~\cite{34p1,34p2} Thus, in order to realize the flexoelectric effect of a solid material, one either uses the limited loading forms (bending or twisting) or fabricates special geometries to bear particular loading forms~\cite{35}. Moreover, currently, the flexoelectric effect of a solid material is relatively low in polymers compared to ceramics~\cite{36}. Realizing a high flexoelectric effect under general loading forms will advance the development of flexoelectricity and broaden its applications.

In this work, we develop a predictive model for the overall flexoelectric effect of a porous material with open pores and ligaments. Because the porous structure is composed of a great amount of randomly oriented ligaments, any kind of loading form at the macroscopic scale will result in a considerable strain gradient in the micro-scale ligaments, thus producing an ultrahigh and omni-directional  weight and deformability specific flexoelectric output. The dependence of the flexoelectric output on the structural properties of the porous material and the external applied load is obtained. Then, to verify the theory, we fabricate the porous materials of polydimethylsiloxane (PDMS) and polyvinylidene fluoride (PVDF), and measure their flexoelectric response. The theoretical results agree well with the experimental data. When a cubic porous PDMS is compressed to a 25$\%$ macroscopic strain, it exhibits ca. 100 times larger  weight and deformability specified flexoelectric response than a solid truncated pyramid. Finally, taking the advantages of large deformation and high flexoelectric sensitivity of the polymer, we apply the model to a bending sensor, and demonstrate its robust functions. Since porous structures can be fabricated with a wide range of materials, such as ceramics~\cite{37}, polymers~\cite{22}, metals~\cite{38} and their corresponding composites~\cite{39,40}, the theoretical and technological approaches can be applied to other materials.

\section{Model and mechanism of flexoelectricity of porous material}

Figure 1 schematically illustrates the bending deformation of a typical internal ligament when a porous material is subjected to tension, compression, bending, and torsion. Non-zero net dipole moments (and hence polarization) will be induced due to the relative displacement between the centers of the positive and negative charges. The induced flexoelectric polarization $P_l$  of the ligament is given by $P_l=\mu_{{ijkl}}({ \partial\varepsilon_{{ij}}}/{\partial x_{k}})$, where $\mu_{ijkl}$ is the flexoelectric coefficient;  $\varepsilon_{{ij}}$ is the elastic strain;  and $x_{k}$ is a coordinate~\cite{41}. The repeated indexes obey the summation convention.

\begin{figure}[h]
  \centering
  \includegraphics[width=10cm]{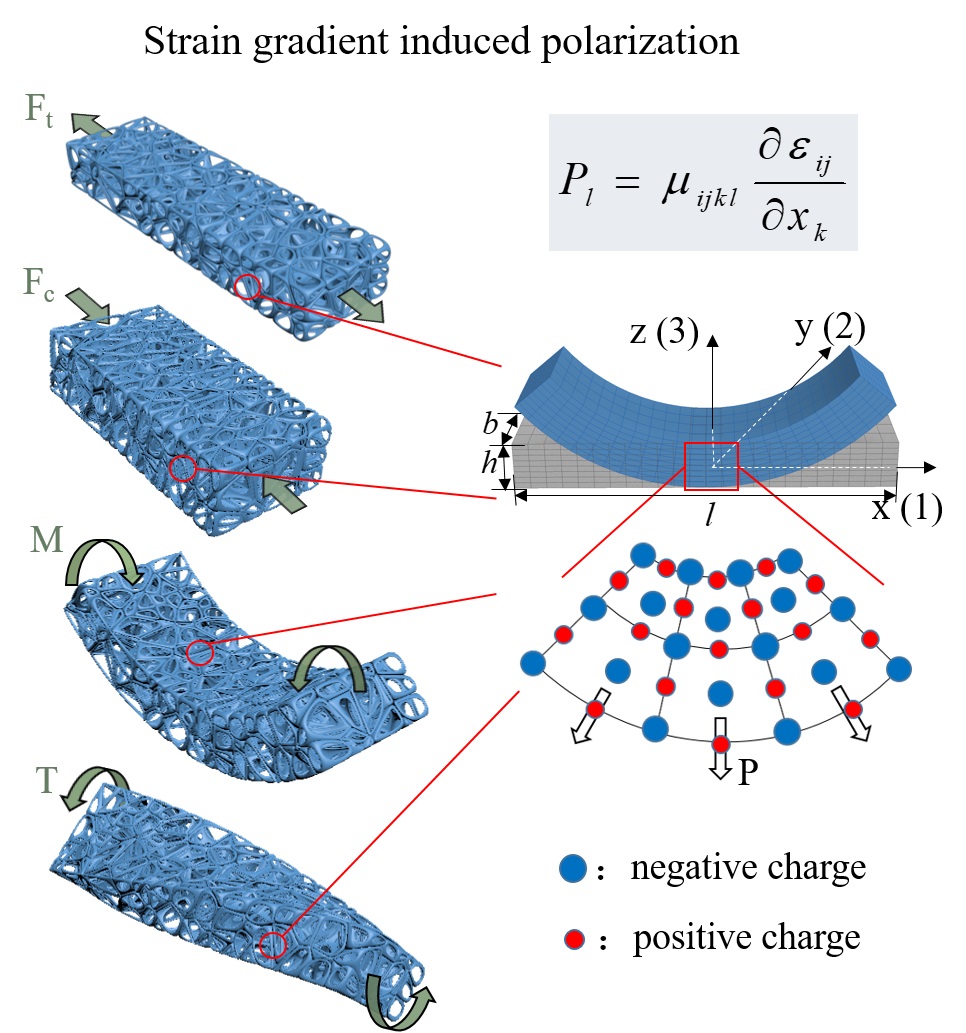}\\
  \caption{Schematic illustration of the deformation of a porous material and the induced flexoelectricity. The porous material under various loading forms such as tension, compression, bending and torsion (left) and the strain gradient induced polarization in the bending deformation of an internal ligament (right).}\label{Fig:1}
\end{figure}

\subsection{Flexoelectricity under macroscopic compression}
In order to predict the flexoelectric response of the porous structure, we propose a theoretical model based on the Gibson-Ashby unit~\cite{42} as shown in Fig.~2(a), and a macroscopic  porous structure, shown in Fig. 2(b), is constructed by the unit cells. The basic element of the unit cell is a beam representing the ligament between two nodes in the porous cell unit.  The unit contains three kinds of beams, namely, $\mathrm{b}_1$ (the beam $\mathrm{b}_1$ shown in Fig. 2(a) is half of the whole beam), $\mathrm{b}_2$, and $\mathrm{b}_3$. These beams have the  same length $l$ and thickness $\delta$.  The Timoshenko beam theory~\cite{43} will be utilized to predict the deformation of the beams, since the aspect ratio of the ligaments of most porous materials is not very high, typically less than 10. To consider the deformation, and thus the flexoelectric polarization of the ligaments more comprehensively, two models are proposed in this work. In the first model, denoted by $\mathrm{M}_1$, the compression of the beams $\mathrm{b}_1$ and $\mathrm{b}_2$ and the bending of beam $\mathrm{b}_3$ are considered. In the second model, denoted by $\mathrm{M}_2$, the bending of the vertical beams $\mathrm{b}_1$ and $\mathrm{b}_2$, along with the compression of the beams $\mathrm{b}_1$ and $\mathrm{b}_2$ and the bending of beam $\mathrm{b}_3$ considered in $\mathrm{M}_1$ are taken into account. Thus, $\mathrm{M}_1$ does not include the bending effect of the vertical beams, and is more suitable for porous materials with small pores, whereas $\mathrm{M}_2$ is more suitable for large pores. We consider a macroscopic cuboid with length $2L$, width $2L$ and height $L$. The number of layers of the unit cells in the vertical direction is $n$. The geometric parameters obey
\begin{equation}\label{9}
n(2l+2\delta)=L.
\end{equation}
We consider a macroscopic displacement applied on the cuboid
\begin{equation}\label{8}
W(t)=\frac{\lambda_\mathrm{{pp}}}{2}-\frac{\lambda_\mathrm{{pp}}}{2} \cos{ 2\pi ft},
\end{equation}
where $\lambda_\mathrm{{pp}}$ is the amplitude of the displacement; $f$ is the frequency; and $t$ represents time.

\begin{figure}[h]
  \centering
  \includegraphics[width=11cm]{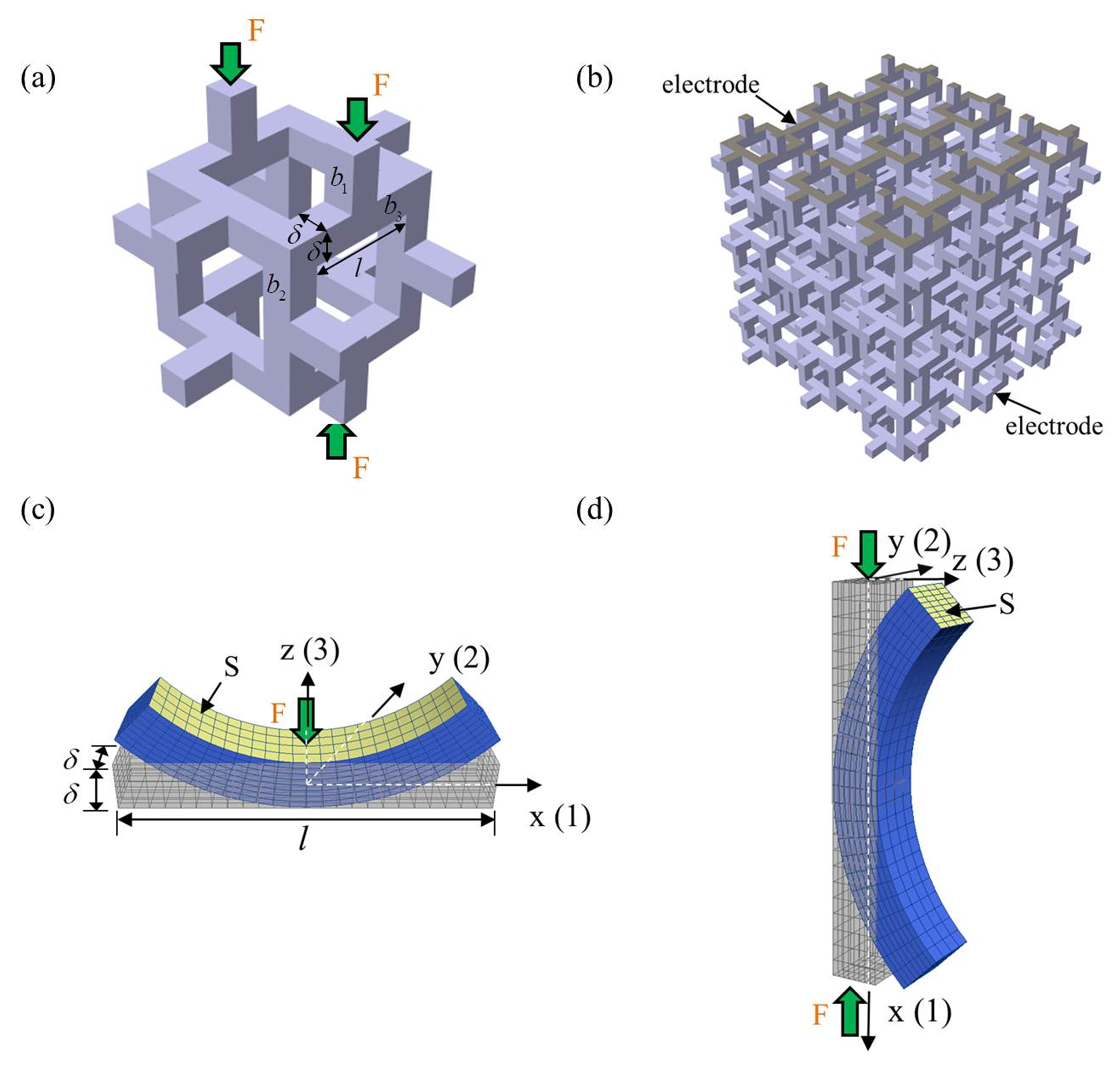}\\
  \caption{Theoretical model of porous material. (a) The Gibson-Ashby unit cell. (b) Macroscopic porous cube with arrays of unit cells. (c) Three-point bending beam element. (d) Compressive beam element.}\label{Fig:2}
\end{figure}

Then, we calculate the deformation of the unit cell under the macroscopic displacement. We start the analysis with an applied concentrated force $F$ at the end of beam $\mathrm{b}_1$, as shown in Fig. 2(a).
In $\mathrm{M}_1$, the deformations of the beams $\mathrm{b}_1$ and $\mathrm{b}_2$ are the shortening in the $x$-direction caused by axial compression, which are denoted as $\triangle d_1$ and $\triangle d_2$, respectively,
\begin{equation}\label{2}
\triangle d_1=\frac{Fl}{EA},
\end{equation}
\begin{equation}\label{3}
\triangle d_2=\frac{Fl}{2EA},
\end{equation}
where $E$ and $A$ are the Young's modulus and the cross-sectional area of the beam, respectively.   The flexoelectric polarization in model $\mathrm{M}_1$ is contributed mainly by the bending of beam $\mathrm{b}_3$.   For the bending deformation of beam $\mathrm{b}_3$ shown in Fig. 2(c), the primary strain is the bending strain $\varepsilon_{11}=-\frac{M(x)z}{EI}$ in the $x$-direction, where $M(x)$ is the bending moment, and $I={\delta^4}/{12}$  is the moment of inertia of the cross-section of the beam. Thus, the polarization vector  is
\begin{equation}\label{pb3v}
{\mathbf P}=\varepsilon_{11,1}(\mu_{1111}{\mathbf e}_{1}+\mu_{1112}{\mathbf e}_{2}+\mu_{1113}{\mathbf e}_{3})+\varepsilon_{11,3}(\mu_{1131}{\mathbf e}_{1}+\mu_{1132}{\mathbf e}_{2}+\mu_{1133}{\mathbf e}_{3}),
\end{equation}
where $\varepsilon_{11,1}$ and $\varepsilon_{11,3}$ are the gradients of the strain $\varepsilon_{11}$ with respect to the coordinates
$x$ and $z$, respectively, and ${\mathbf e}_{i}\ (i=1,2,3)$ represent the unit base vectors of the Cartesian coordinate system. Theoretically,
under the three-point bending deformation shown in Fig.~2(c), the bending moment $M(x)=\frac{F}{2}(\frac{l}{2}-|x|)$ is a function of the  coordinate $x$, and thus
the strain gradient $\varepsilon_{11,1}$ is generally not zero. However, compared to the gradient $\varepsilon_{11,3}$, the effect of $\varepsilon_{11,1}$ is generally
neglected~\cite{07}. Then, among the three components of the polarization vector ${\mathbf P}$ induced by $\varepsilon_{11,3}$, we only collect
the charges corresponding to $\varepsilon_{11,3}\mu_{1133}{\mathbf e}_{3}$. Thus,
 the magnitude of the effective electric polarization $P_{\mathrm{b}_{3}}$ of beam $\mathrm{b}_3$ due to the flexoelectric effect can be calculated as

\begin{equation}\label{pb3w}
P_{\mathrm{b}_{3}}=\mu_{1133}  \frac{ \frac{F}{2} (\frac{l}{2}-|x|)}{EI},
\end{equation}
where  $-l/2\leq x \leq l/2$ ~(Fig.~2).

The flexoelectric charge can be collected of a single beam $b_{3}$ is $Q_\mathrm{b}$, which is given by
 \begin{equation}\label{pb3wj}
 Q_{\mathrm{b}}=\int P_{\mathrm{b}_{3}}\mathrm{d}S=\delta \int_{-l/2}^{l/2}\mu _{1133}\frac{\frac{F}{2}(\frac{l}{2}-|x|)}{EI}\mathrm{d}x=\frac{\mu _{1133}F\delta l^{2}}{8EI},
 \end{equation}

 There are two $\mathrm{b}_3$ beams on each side of the unit cell. Thus, for the unit cell in model $\mathrm{M}_1$, the flexoelectric charge is
\begin{equation}\label{10}
Q_1=2Q_\mathrm{b}=\frac{\mu_{1133}F\delta l^2}{4EI}.
\end{equation}
Normally, the electrodes to collect the charges only locate on the most top and bottom surfaces of the porous specimen.  When the electric current is measured by connecting the electrodes on the upper and bottom surfaces to the current preamplifier, one wire of the surfaces  is virtually grounded, as described in the
Supplementary Material,
which means that only the flexoelectric charges  on one surface ($2n\times2n$ cells in total) are collected to generate the current. Thus, the measured flexoelectric polarized charges of the whole porous cuboid can be calculated as
\begin{equation}\label{11}
 Q_{\mathrm{sum1}}=4n^2 Q_1=\frac{\mu_{1133} Fn^2\delta l^2}{EI}.
\end{equation}
Then, the relation between $F$ in Eq. (8) and the known applied displacement load $W(t)$ in Eq. (2) needs to be established.

The displacement of the unit cell along the vertical direction caused by the bending of beam $\mathrm{b}_3$ is denoted as $\triangle d_3$.
The total displacement of the unit cell in the vertical direction is $\triangle d_{\mathrm{M}_1}=\triangle d_1+\triangle d_2+\triangle d_3$ and obeys
\begin{equation}\label{13}
n\cdot\triangle d_{\mathrm{M}_1}=W(t).
\end{equation}
Then, we calculate $\triangle d_3$. Half of the beam $\mathrm{b}_3$ in Fig. 2(c) is considered as a cantilever beam because of the symmetry. Based on the Timoshenko beam theory, the three components $u_\mathrm{x}$, $u_\mathrm{y}$, $u_\mathrm{z}$ of the displacement of the cantilever beam are
\begin{eqnarray*}
u_{\mathrm{x}} (x,y,z)=-z\varphi(x),\;
u_{\mathrm{y}} (x,y,z)=0,\;
u_{\mathrm{z}} (x,y)=w_{\mathrm{z}}(x),
\end{eqnarray*}
where $(x, y,z)$ represents the coordinates of a point in the beam; $\varphi$ represents the bending angle of the beam's neutral plane; and $w_{\mathrm{z}}$ is the displacement of the neutral plane in the z-direction. The governing equations of the Timoshenko beam are
\begin{eqnarray*}
\frac{\mathrm{d}^2}{\mathrm{d}x^2}\left(EI_\mathrm{y}  \frac{\mathrm{d}\varphi}{\mathrm{d} x}\right)=q(x,t),\;\;
	\frac{\mathrm{d}w_{\mathrm{z}}}{\mathrm{d}x}=\varphi-\frac{1}{\kappa AG}  \frac{\mathrm{d}}{\mathrm{d}x} \left(EI_\mathrm{y}  \frac{\mathrm{d}\varphi}{ \mathrm{d}x}\right),
\end{eqnarray*}
where $q$ is the distributed load (here, $q$=0); $G$ is the shear modulus; $A$ is the cross-sectional area; $\kappa={10(1+v)}/({12+11v})$ is the shear correction factor; and $v$ is the Poisson ratio. The boundary conditions are
\begin{eqnarray*}
&w_\mathrm{z} |_{\mathrm{x}=0} =\varphi|_{\mathrm{x}=0} =0,\;\;
M_{\mathrm{x}= \frac{l}{2}}=EI_\mathrm{y} \left(\frac{\mathrm{d}\varphi}{ \mathrm{d}x}\right)_{\mathrm{x}= \frac{l}{2}}=0,\;\;\\
&F_{s_{\mathrm{x}= \frac{l}{2}}}=\kappa GA\left(\frac{\mathrm{d}w}{\mathrm{d}x}-\varphi\right)_{\mathrm{x}= \frac{l}{2}}=\frac{F}{2},
\end{eqnarray*}
where $M$ is the bending moment. The solutions of $\varphi$ and $w_\mathrm{z}$ are
\begin{eqnarray*}
\varphi(x)=\frac{Fx(x-l)}{4EI},\;\;
w_\mathrm{z}(x)=\frac{Fx^2(2x-3l)}{24EI}-\frac{Fx}{2\kappa AG}.
\end{eqnarray*}
Thus, we have
\begin{equation}\label{14}
\triangle d_3=2w_\mathrm{z}(\frac{l}{2})=\frac{Fl^3}{24EI}+\frac{Fl}{2\kappa AG}.
\end{equation}
Combining Eqs. (\ref{2})-(\ref{3}) with Eqs. (\ref{11})-(\ref{14}), the total flexoelectric charge of the whole porous cuboid based on model $\mathrm{M}_1$ can be calculated as
\begin{equation}\label{15}
 Q_\mathrm{sum1} =\mu_{1133}\frac {6Ll\kappa G\delta}{(l+\delta)}\cdot\frac{\lambda_\mathrm{{pp}} -\lambda_\mathrm{{pp}} \cos{ 2\pi ft}}{3\kappa G\delta^2+\kappa Gl^2+E\delta^2 }
\end{equation}
and the corresponding total current is
\begin{equation}\label{16w}
I=\frac{\mathrm{d} Q_\mathrm{sum1} }{\mathrm{d}t}=\mu_{1133}\frac {12Ll\kappa G\delta}{(l+\delta)}\cdot\frac{\lambda_\mathrm{{pp}} \pi f \sin{ 2\pi ft}}{3\kappa G\delta^2+\kappa Gl^2+E\delta^2}.
\end{equation}
So far, the dependence of the polarized current $I$ from flexoelectricity on the applied displacement load $W(t)$ based on model $\mathrm{M}_1$ is obtained.

In model $\mathrm{M}_1$, only the axial compression of the vertical beams is taken into account, which has nearly no contribution to the flexoelectric effect. In order to describe the deformation comprehensively, the bending of the vertical beams $\mathrm{b}_1$ and $\mathrm{b}_2$ is taken into account in  model $\mathrm{M}_2$. Considering that the ligaments in real irregular porous materials are hardly perfectly straight, we assign a relative pre-deflection $\delta_0' = {\delta_0}/{\delta}=0.2$ at the midpoint of the vertical beam.  The method to determine the value of pre-deflection of the porous material is given in the Supplementary Material. Then, the displacements along the vertical direction of the unit cell caused by the bending of beam $\mathrm{b}_1$ and $\mathrm{b}_2$ are \cite{44}
\begin{equation}\label{16}
\triangle d_{1\mathrm{v}}=\frac{\pi^2\delta_0^2}{2l}\frac{F(P_\mathrm{E}-0.5F)}{(P_\mathrm{E}-F)^2},
\end{equation}
\begin{equation}\label{17}
\triangle d_{2\mathrm{v}}=\frac{\pi^2\delta_0^2}{2l}\cdot\frac{F(4P_\mathrm{E}-F)}{2(2P_\mathrm{E}-F)^2},
\end{equation}
where $P_\mathrm{E}={\pi^2EI}/{l^2}$.
The total vertical displacement of the unit cell can be calculated as
\begin{equation}\label{18}
\triangle d_{\mathrm{M}_2 }=\triangle d_1+\triangle d_2+\triangle d_3+\triangle d_{1\mathrm{v}} +\triangle d_{2\mathrm{v}} ,
\end{equation}
where $\triangle d_1$, $\triangle d_2$ and $\triangle d_3$ can be obtained using Eqs. (\ref{2}),  (\ref{3}) and (\ref{14}), respectively.

The flexoelectric response of the unit cell in model $\mathrm{M}_2$ consists of both the contribution of the bending of beam $\mathrm{b}_3$ considered in $\mathrm{M}_1$ in Eq. (\ref{10})), and also the contribution of beams $\mathrm{b}_1$ (denoted as $Q_2$). However, it is noted that the length and thickness of the ligaments in $\mathrm{M}_2$ are different from those in $\mathrm{M}_1$.
The contribution of beam $\mathrm{b}_2$ is not taken into account since it can be hardly collected by the electrode at the top or bottom layer.

The vertical beam $\mathrm{b}_1$ is modelled by a hinged-supported slender beam subjected to an axial force $F$ at its ends (Fig.~2(d)).
 Its deflection can be taken as a sine curve
\begin{equation}\label{5}
w_\mathrm{c} (x)=w_\mathrm{c}(\frac{l'}{2})\sin{\frac{\pi}{l'}x},
\end{equation}
where $l'=l- \triangle d_{1}-\triangle d_{1\mathrm{v}}$ is the chord length after deformation, and $ w_\mathrm{c}({l'}/{2})=({1}/{2}) \sqrt{l^2-l'^2}$ is the deflection at the midpoint of the beam (the arc length is approximated by the chord length).  Then, applying the
expression (\ref{pb3v}) of the polarization vector to beam $\mathrm{b}_1$ in Fig.~2(d), the electric polarization
 is

\begin{equation}\label{6}
P_{\mathrm{b}_{1}}=\mu_{1131}\varepsilon_{11,3}=\mu_{1131}\frac{Fw_\mathrm{c}(\frac{l'}{2})\sin{\frac{\pi}{l'}x}}{EI}.
\end{equation}

The electric charge can be collected by a single beam $\mathrm{b}_1$ is $Q_\mathrm{c}$, which is given by
\begin{equation}\label{7}
Q_\mathrm{c}=\overline{P_{\mathrm{b}_{1}}}S=\mu_{1131} \frac{F\delta^2\sqrt{l^2-l'^2}}{\pi EI},
\end{equation}
where $\overline{P_{\mathrm{b}_{1}}}$ is the average value of $P_{\mathrm{b}_{1}}$ along $x$ direction and $S$ is a constant, $S$=$\delta^2$.
Thus, the flexoelectric response of the unit cell in model $\mathrm{M}_2$ is
\begin{equation}\label{19}
 Q_2=2\mu_{1131}\frac {F\delta^2 \sqrt{l^2-l'^2 }}{\pi EI}.
\end{equation}

Thus, the measured flexoelectric charge of the whole porous material based on model $\mathrm{M}_2$ can be calculated as
\begin{equation}\label{20}
 Q_\mathrm{sum2} =4n^2 (Q_1+Q_2),
\end{equation}
and the corresponding current is
\begin{equation}\label{21}
I=\frac{\mathrm{d}Q_\mathrm{sum2} }{\mathrm{d}t}.
\end{equation}
Similarly to Eq. (\ref{13}), $F$ can be related to $W(t)$ through
\begin{equation}\label{22}
n\cdot\triangle d_{\mathrm{M}_2}=W(t).
\end{equation}
Finally, combining Eqs. (\ref{2}), (\ref{3}), (\ref{14}) with Eqs. (\ref{16})-(\ref{22}), the dependence of the flexoelectric polarized current $I$ on the applied displacement  $W(t)$ based on model $\mathrm{M}_2$ is obtained.

Based on the theoretical models $\mathrm{M}_1$ and $\mathrm{M}_2$, the dependence of the flexoelectric current on different parameters including the pore size, ligament thickness, Poisson ratio, and macroscopic displacement are shown in Fig. 3. The used parameters are as follows: the macroscopic length, width and height of the porous material are 20 $\mathrm{mm}$, 20 $\mathrm{mm}$ and 10 $\mathrm{mm}$, respectively; the pore diameter and ligament thickness are 180 $\mathrm{\mu m}$ and 50 $\mathrm{\mu m}$, respectively; the macroscopic compression displacement, frequency and Poisson ratio of the material are 2.5 $\mathrm{mm}$, 3 Hz, and 0.49, respectively. The flexoelectric current of the porous structure increases significantly with the decrease of the pore diameter and the increase of the ligament thickness for both models. Moreover, the flexoelectric current increases linearly with the decrease of the Poisson ratio and the increase of the macroscopic compression displacement. The flexoelectric output current of model $\mathrm{M}_2$ is always higher than that of model $\mathrm{M}_1$, which is reasonable since all the possible bending is considered in model $\mathrm{M}_2$.

\begin{figure}[h]
  \centering
  \includegraphics[width=12cm]{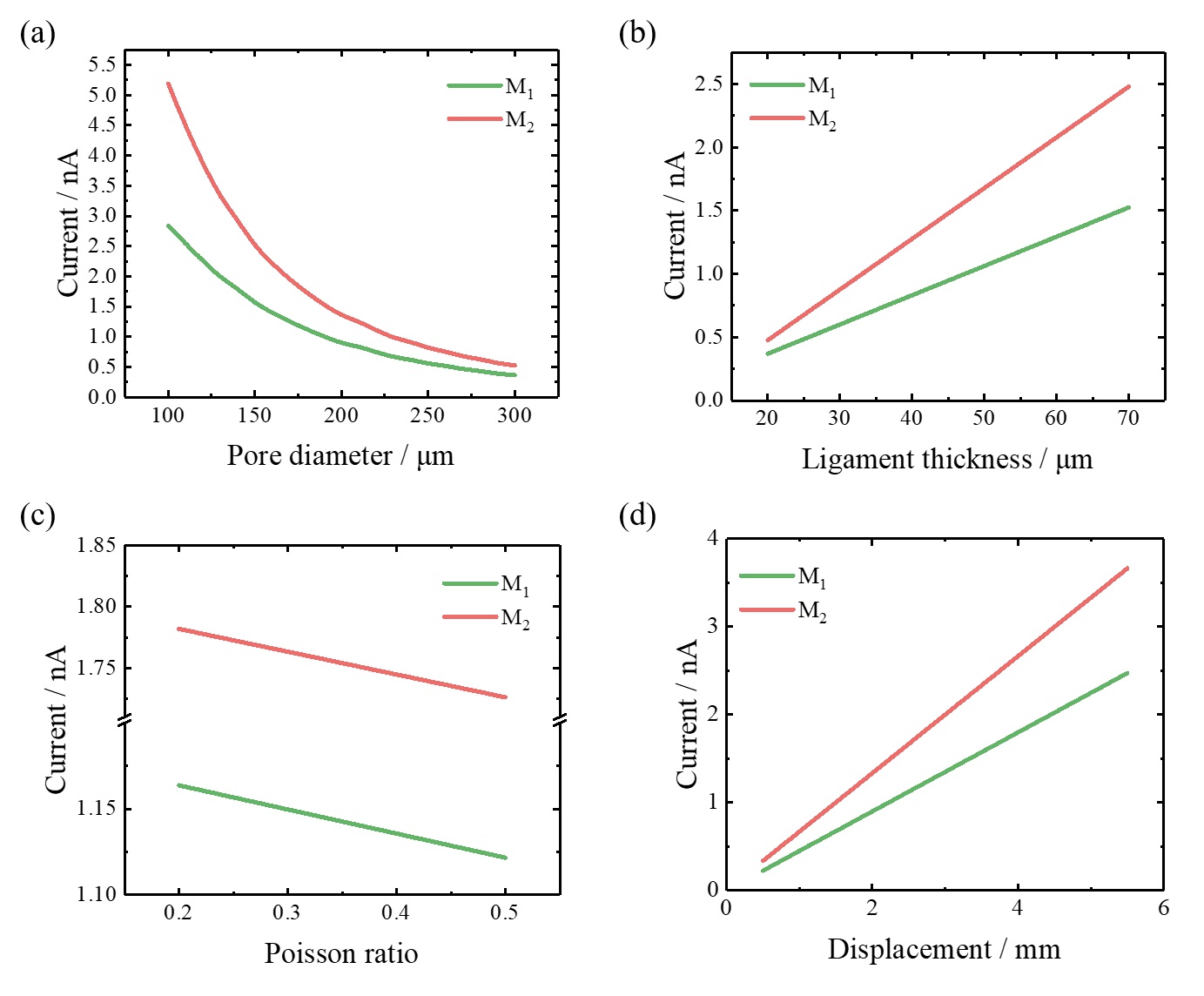}\\
  \caption{  Dependence of flexoelectric current on (a) pore diameter ($ \textit{l}=180\,\mathrm{\mu m},\textit{f}=3 \,\mathrm{Hz}, \lambda_{ \mathrm{pp}}=2.5\, \mathrm{mm},\nu=0.49  $), (b) ligament  thickness ($\delta =50\,\mathrm{\mu m},\textit{f}=3 \,\mathrm{Hz}, \lambda_{ \mathrm{pp}}=2.5\, \mathrm{mm},\nu=0.49  $), (c) Poisson ratio ($\textit{l}=180\,\mathrm{\mu m},\delta =50\,\mathrm{\mu m}, \textit{f}=3 \,\mathrm{Hz}, \lambda_{ \mathrm{pp}}=2.5\, \mathrm{mm}$) and  (d) macroscopic compression displacement ($\textit{l}=180\,\mathrm{\mu m},\delta =50\,\mathrm{\mu m}, \textit{f}=3 \,\mathrm{Hz},\nu=0.49  $).} \label{Fig:3}
\end{figure}

\subsection{A dimensional analysis}

The effective current of $\mathrm{M}_1$ can be given by a simple formula in Eq.~(\ref{16w}), but the result of $\mathrm{M}_2$ cannot be expressed in a simple one. Here, we conduct a dimensional analysis to reveal the dependence of the effective current of a general porous material on the major governing parameters.
For a given porous material, the effective current is a function of the geometric and material parameters as follows:
\begin{equation}\label{22w1}
I=J(\mu,l,\delta,\lambda_\mathrm{{pp}},f,L,\nu).
\end{equation}
Here, the meanings of the variables are the same as those in the above derivations. By a dimensional analysis, this formula can be expressed as
\begin{equation}\label{22w2}
I=\mu l fJ^*(\frac{\delta}{l},\frac{\lambda_\mathrm{{pp}}}{l},\frac{L}{l},\nu).
\end{equation}
The parameters $L/l$ is related to the total layers of the unit cells in a macroscopic structure. However, the collected current is proportional to the layers where the charges are collected (usually, by coating the surfaces of the macroscopic structure), but not to the total layers. Thus, it is reasonable to write Eq.~(\ref{22w2}) approximately as
\begin{equation}\label{22w3}
I=n^*\mu l fJ^*(\frac{\delta}{l},\frac{\lambda_\mathrm{{pp}}}{l},\nu),
\end{equation}
where $n^*$ denotes the total layers where the charges are collected. $\delta/l$ is a small quantity (eg., about 0.1), and when $\delta$ and
 $\lambda_\mathrm{{pp}}/{l}$ are zero, there will be no current. Thus, by the Taylor expansion, we can approximate Eq.~(\ref{22w3}) by
 \begin{equation}\label{22w4}
I=n^*\mu l f \frac{\delta}{l}\frac{\lambda_\mathrm{{pp}}}{l} J^{**}(\nu)=n^*\mu \delta f \frac{\lambda_\mathrm{{pp}}}{l} J^{**}(\nu).
\end{equation}
This formula clearly shows the linear dependence of the effective current on the thickness $\delta$, the applied macroscopic displacement
$\lambda_{pp}$, and the frequency $f$, as well as the linear dependence on the inverse of the pore size $l$.

From Eq.~(\ref{22w4}), we can get the ratio of the effective current  to the relative density  of the porous material
 \begin{equation}\label{22w5}
I^{*}\equiv I/\rho^*=\alpha n^*\mu f \lambda_\mathrm{{pp}}\rho^{*-1/2},
\end{equation}
where $\alpha$ is a non-dimensional parameter related to the geometry of the microstructure, and $\rho^{*}=\rho_\mathrm{{porous}}/\rho_\mathrm{{solid}}$ is the
relative density.  Thus, $I^*$ scales with $\rho^{*-1/2}$, demonstrating the enhancement of the  weight specific
flexoelectric response due to the light-weight. In order to manifest the enhancing effect of the porous structure, considering that
$\rho^{*}=\rho_\mathrm{{porous}}/\rho_\mathrm{{solid}} \propto \frac{\delta^2 l}{l^{2}\delta}=\delta/l$, we rewrite
Eq.~(\ref{22w5}) as
\begin{equation}\label{22w56}
I^{*}\propto\alpha \mu f \lambda_\mathrm{{pp}}n^*\left(\frac{l}{\delta}\right)^{1/2}
\end{equation}
This formula manifests the enhancing effect of the porous structure where, generally, $l/\delta>>1$; it also shows the benefit of collecting charges from more ligaments.

\subsection{Flexoelectricity under macroscopic bending}

 Now, we consider a macroscopic beam of a porous microstructure under macroscopic bending deformation. As shown in Fig. 4, the macroscopic length, width and height of the beam are $l_0$, 0.25$l_0$, and $h$, respectively. The beam has the same microstructure as the aforementioned models $\mathrm{M}_1$ and $\mathrm{M}_2$ as shown in Fig.2(a). When the beam is subjected to bending deformation, its unit cell in the top layer is under the force as shown in Fig. 4(c). The flexoelectric charges are mainly from the bending of beam $\mathrm{b}_{2}$.
Similarly to Eqs. (\ref{5}) and (\ref{6}), the electric polarization $P_3$ due to the flexoelectric effect can be calculated as
\begin{equation}\label{23}
P_3=\mu_{1133}\frac {F w_2(\frac{l_2'}{2})\sin \frac{\pi}{l_2'}x} {2EI},
\end{equation}
where $l_2'=l-\triangle d_2-\triangle d_{2\mathrm{v}}$  is the chord length of beam $\mathrm{b}_{2}$ after deformation. The horizontal displacements of beam $\mathrm{b}_{2}$ due to axial compression and bending are the same as $\triangle d_2$ and $\triangle d_{2\mathrm{v}}$ described in Eqs. (\ref{3}) and (\ref{17}), and $w_2 ({l_2'}/{2})=({1}/{2}) \sqrt{l^2-l_2 '^2 }$ is the deflection at the midpoint of beam $b_3$. The induced charge of the unit cell that can be collected is
\begin{equation}\label{24}
Q_3=2\int P_3  \mathrm{d}S =2\delta\int_{0}^{l_2'} P_3  \mathrm{d}x=  \frac{\mu_{1133} Fl_2'\delta \sqrt{l^2-l_2 '^2 }}{\pi EI  }.
\end{equation}
 Again, one layer of unit cell at the top of the sample is considered. The number of the layers of unit cells along the axial direction of the macroscopic beam is
\begin{equation}\label{30}
m=\frac{l_0}{2(l+\delta)}.
\end{equation}
Then the induced charge and corresponding current of the whole porous structured beam are
\begin{equation}\label{25}
Q_\mathrm{{sum3}}=\frac{m^2}{4 }Q_3,
\end{equation}
\begin{equation}\label{26}
I=\frac{\mathrm{d}Q_\mathrm{{sum3}}}{\mathrm{d}t}.
\end{equation}
$F$ can be related to the applied bending angle $\theta(t)$ as follows:
\begin{equation}\label{27}
\theta(t)=\frac{\theta_{\mathrm{pp}}}{2}-\frac{\theta_{\mathrm{pp}}}{2}\cos2\pi ft,
\end{equation}
where $\theta_{\mathrm{pp}}$  is the peak to peak angle.
As shown in Fig. 4(b), the deformation of the porous beam satisfies the following equation:
\begin{equation}\label{28}
R=\frac{l_0-\triangle d_0}{\theta}=\frac{l_0+\triangle d_0}{\theta}-h,
\end{equation}
where $R$, $\triangle d_0$ are the radius and the length decrease of the top surface, respectively.
Thus
\begin{equation}\label{29}
\triangle d_0=\frac{h\theta}{2}.
\end{equation}

The horizontal displacement of the unit cell, ${\triangle d_0}/{m}$, can be calculated as
\begin{equation}\label{31}
\frac{\triangle d_0}{m}=\triangle d_1+\triangle d_2+\triangle d_3+\triangle d_{1\mathrm{v}} +\triangle d_{2\mathrm{v}}.
\end{equation}
Therefore, combining Eqs. (\ref{2}), (\ref{3}), (\ref{14}), (\ref{16}), (\ref{17}), and Eqs. (\ref{24})-(\ref{31}), the dependence of the flexoelectric polarized current $I$ on the applied deformation $\theta(t)$ can be obtained.

\begin{figure}[h]
  \centering
  \includegraphics[width=12cm]{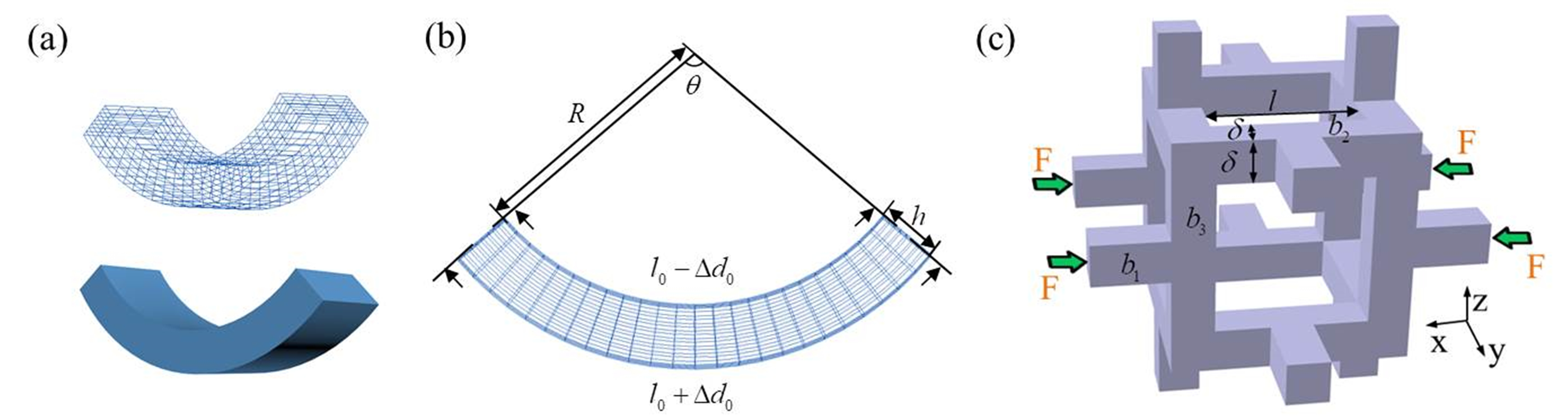}\\
  \caption{ Bending deformation of a porous beam. (a) Macroscopic bending of a porous (top) and solid (bottom) beam. (b) Macroscopic geometric relations of the porous beam. (c) The unit cell of the porous beam.}\label{Fig:4}
\end{figure}

Figure 5 illustrates the dependence of the flexoelectric output current of a macroscopic bending beam with the porous microstructure ($60\times15\times5 \;\mathrm{mm}^3$) on the pore diameter, ligament thickness, Poisson ratio and bending angle. The input parameters are pore diameter, ligament thickness, macroscopic bending angle, bending frequency and Poisson ratio, which are 180 $\mu$m, 50 $\mu$m, $60 ^{\circ}$, 3 Hz and 0.49, respectively.
The flexoelectric current of the porous beam increases significantly with the decrease of the pore diameter and Poison ratio, and the increase of the ligament thickness. These trends are the same as those of model $\mathrm{M}_1$ and $\mathrm{M}_2$. As shown in Fig. 5(d), the flexoelectric output current increases nearly linearly with the bending angle.
\begin{figure}[h]
  \centering
  \includegraphics[width=11cm]{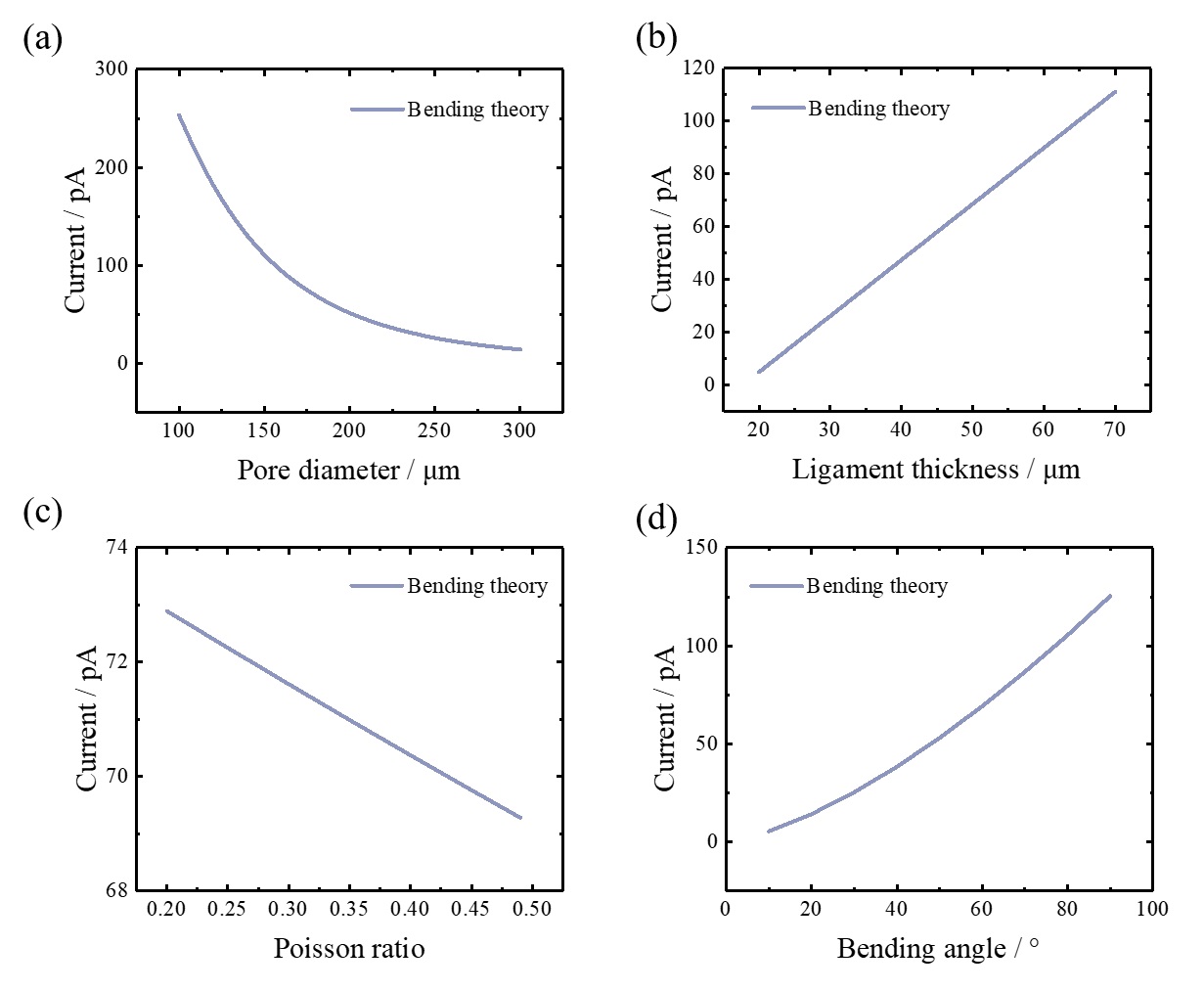}\\
  \caption{ Dependence of flexoelectric output current of a macroscopic bending beam with porous microstructure ($60\times15\times5 \mathrm{mm}^3$) on (a) pore diameter, (b) ligament thickness, (c) Poisson ratio, and (d) bending angle.}\label{Fig:5}
\end{figure}

\section{Verification: porous PDMS and PVDF  and their flexoelectricity}

\subsection{Flexoelectric response of porous PDMS}

Porous PDMS specimens (shown in Fig. 6(a)) were fabricated by the direct templating technique as schematically demonstrated in Supplementary Fig. 1. Unless otherwise indicated, the macro-size of the specimen we used for the measurement is $20\times20\times10 \;\mathrm{mm}^3$. Fig. 6(b) are the SEM images of the porous PDMS specimens, which illustrate that the specimens consist of 3D interconnected structures with open pores and solid ligaments. Here, we measure the flexoelectric properties of two kinds of specimens with different pore sizes, namely, ``small pore'' (Fig. 6(b), left), and ``large pore'' (Fig. 6(b), right). The average pore diameter and ligament thickness of the specimens with small pores are 180 $\mu$m and 50 $\mu$m, respectively, and those of the  specimens with large pores are 400 $\mu$m and 75 $\mu$m, respectively.

\begin{figure}[h]
  \centering
  \includegraphics[width=12cm]{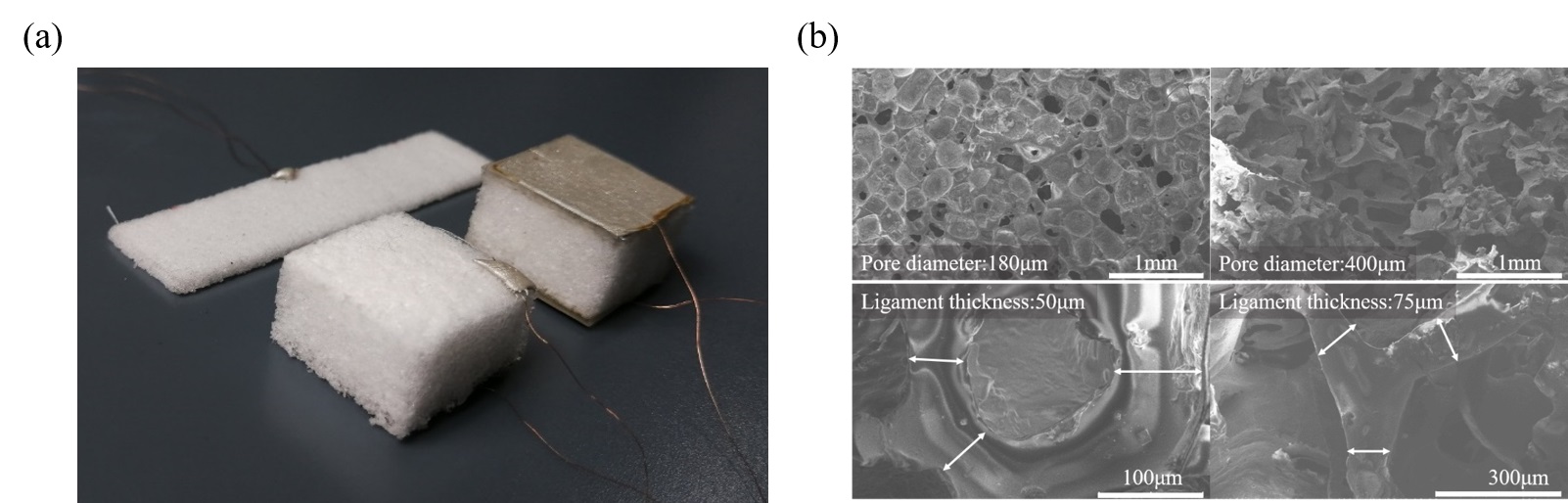}\\
  \caption{ Porous PDMS specimens and their microstructures. (a) Porous PDMS specimens with different macroscopic geometries. (b) SEM images of the 3D interconnected structures with open small pores (left) and large pores (right).}\label{Fig:6}
\end{figure}

In the compression experiment, a dynamic compression displacement in Eq. (2) is applied on the porous sample. As illustrated in Fig. 7(a), the flexoelectric-induced polarization current (left axis) of the PDMS specimens of large pores and the real-time applied displacement (right axis) display remarkable and repeatable signals, and the current frequency (3 Hz) is highly consistent with that of the applied load. Here, unless otherwise indicated, a periodic applied load of 2 Hz is employed in this work, and the top and bottom surfaces of the specimen are covered with silver conductive electrodes to collect the flexoelectric polarized charges.
From Fig. 7(b), one can see that the maximum output current $I_\mathrm{max}$ increases linearly with the increase of the compression strain with both kinds of porous specimens, and  $I_\mathrm{max}$ of the  PDMS specimens with small pores is about twice that of the specimens of large pores. Therefore, as the size of the microstructure decreases, the amplitude of the flexoelectric response increases significantly.

\begin{figure}[h]
  \centering
  \includegraphics[width=12cm]{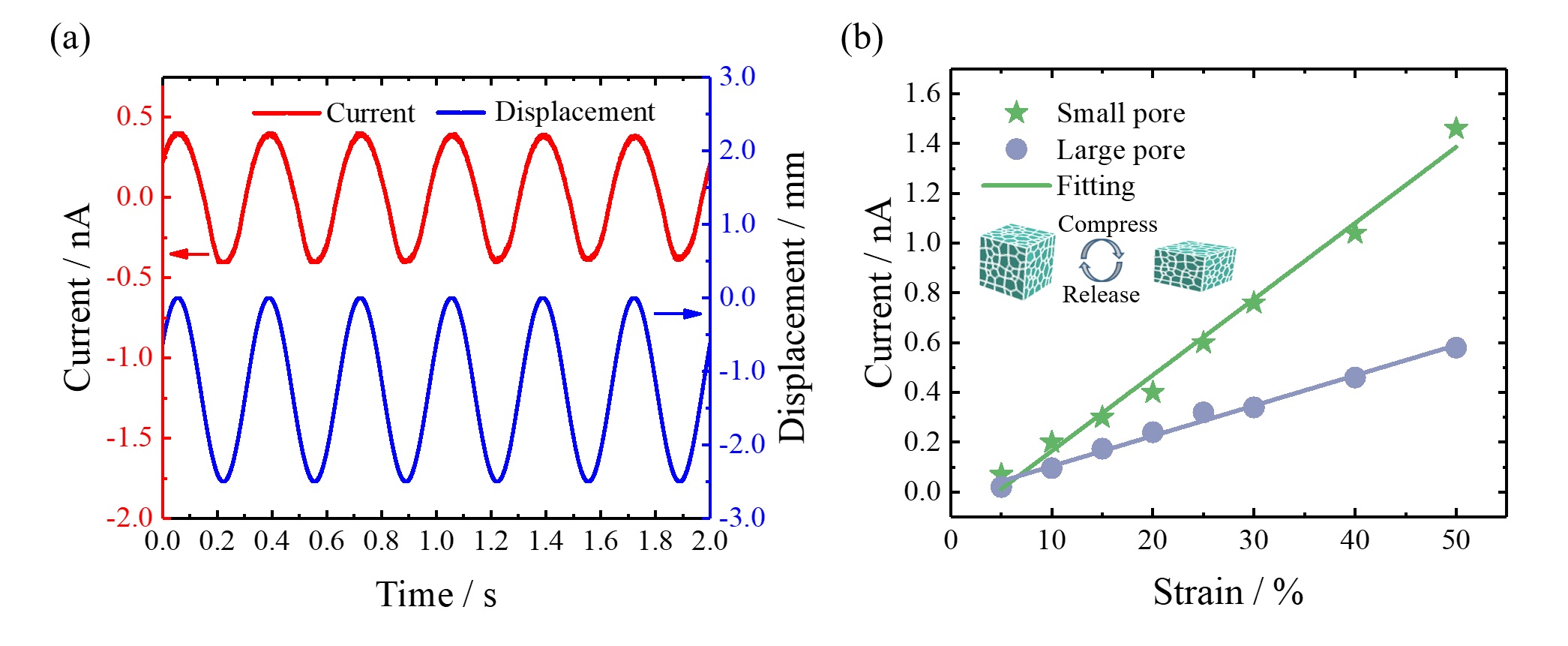}\\
  \caption{Flexoelectric responses of porous PDMS. (a) The flexoelectric-induced current (left axis, red) of large pore PDMS and the real-time applied displacement (right axis, blue) at 3 Hz. (b) The maximum output current-compression strain curves of two kinds of porous PDMS samples.}\label{Fig:7}
\end{figure}

\begin{figure}[h]
  \centering
  \includegraphics[width=12cm]{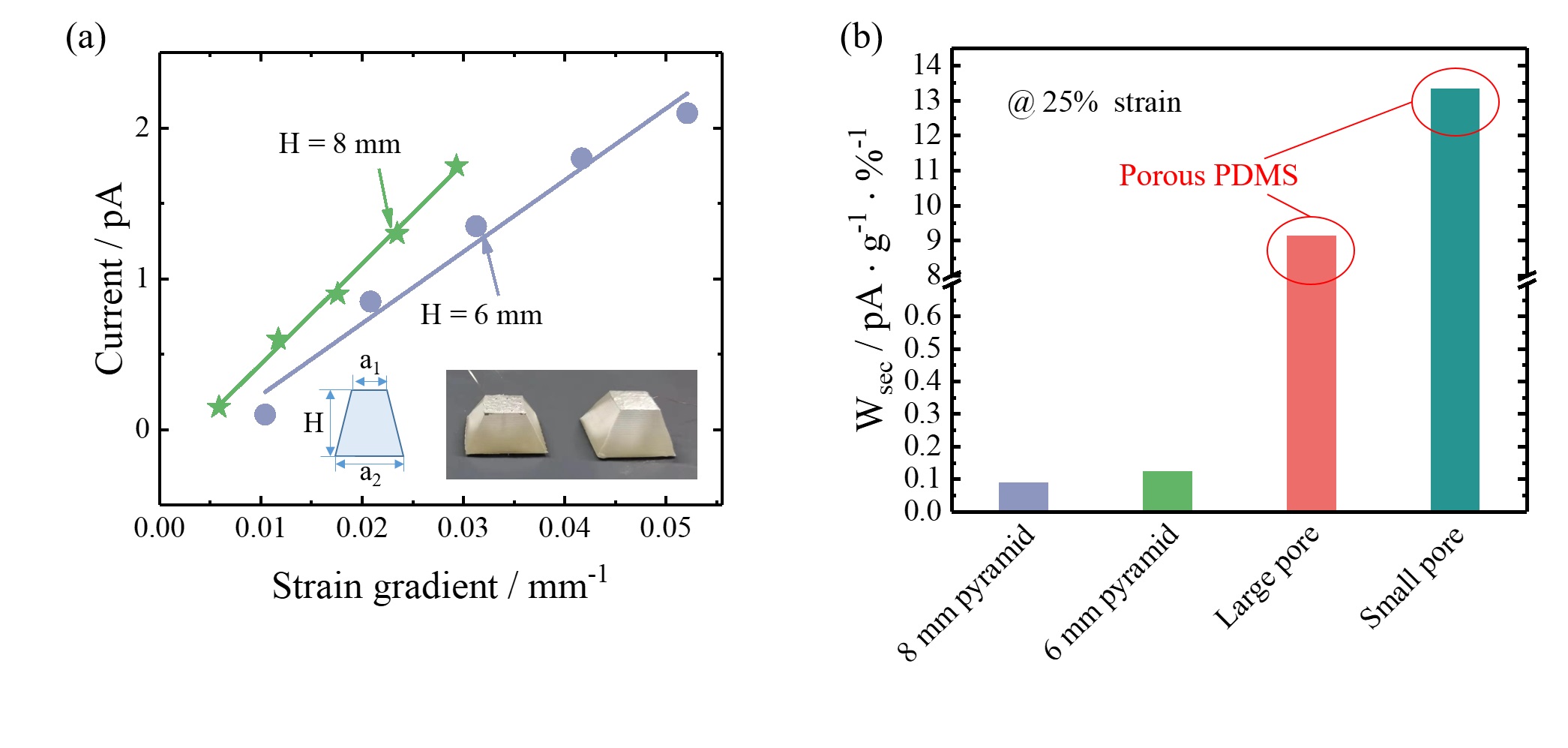}\\
  \caption{Comparison of the  weight and deformability specific effective coefficients $W_\mathrm{sec}$ of porous and solid PDMS. (a) The maximum current-strain gradient curves of two truncated PDMS pyramids with different heights. (b) $W_\mathrm{sec}$ of different specimens under the applied strain  25$\%$. $W_\mathrm{sec}$ of the porous PDMS specimens is two orders of magnitude higher than that of the truncated pyramids.}\label{Fig:8}
\end{figure}

\begin{figure}[h]
  \centering
  \includegraphics[width=12cm]{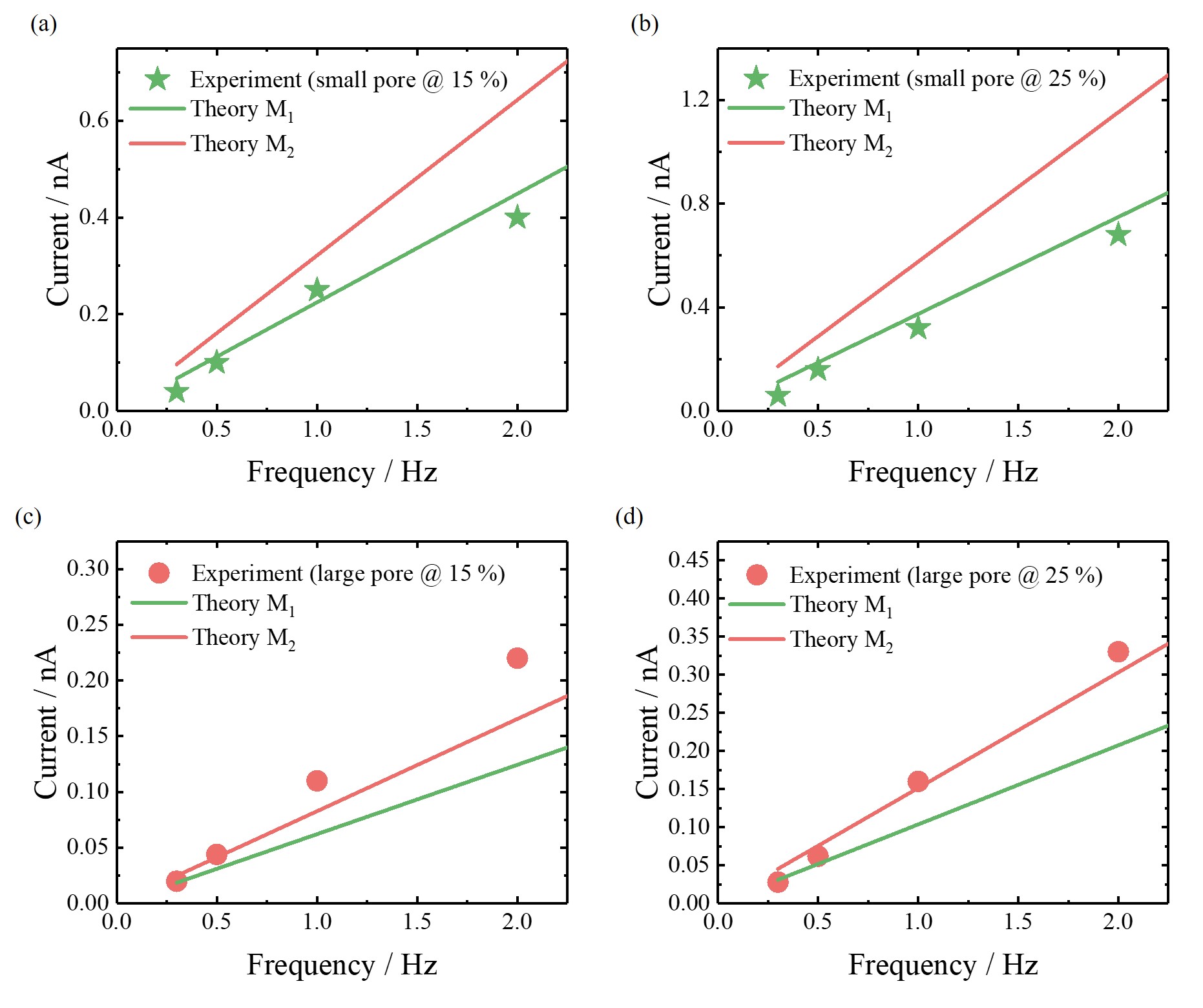}\\
  \caption{Comparison  between the experimental and theoretical flexoelectric responses under different frequencies: PDMS of small pores (a, b) and large pores (c, d) under different frequencies at 15$\%$ and 25$\%$ strains.}\label{Fig:9}
\end{figure}

\begin{figure}[h]
  \centering
  \includegraphics[width=12cm]{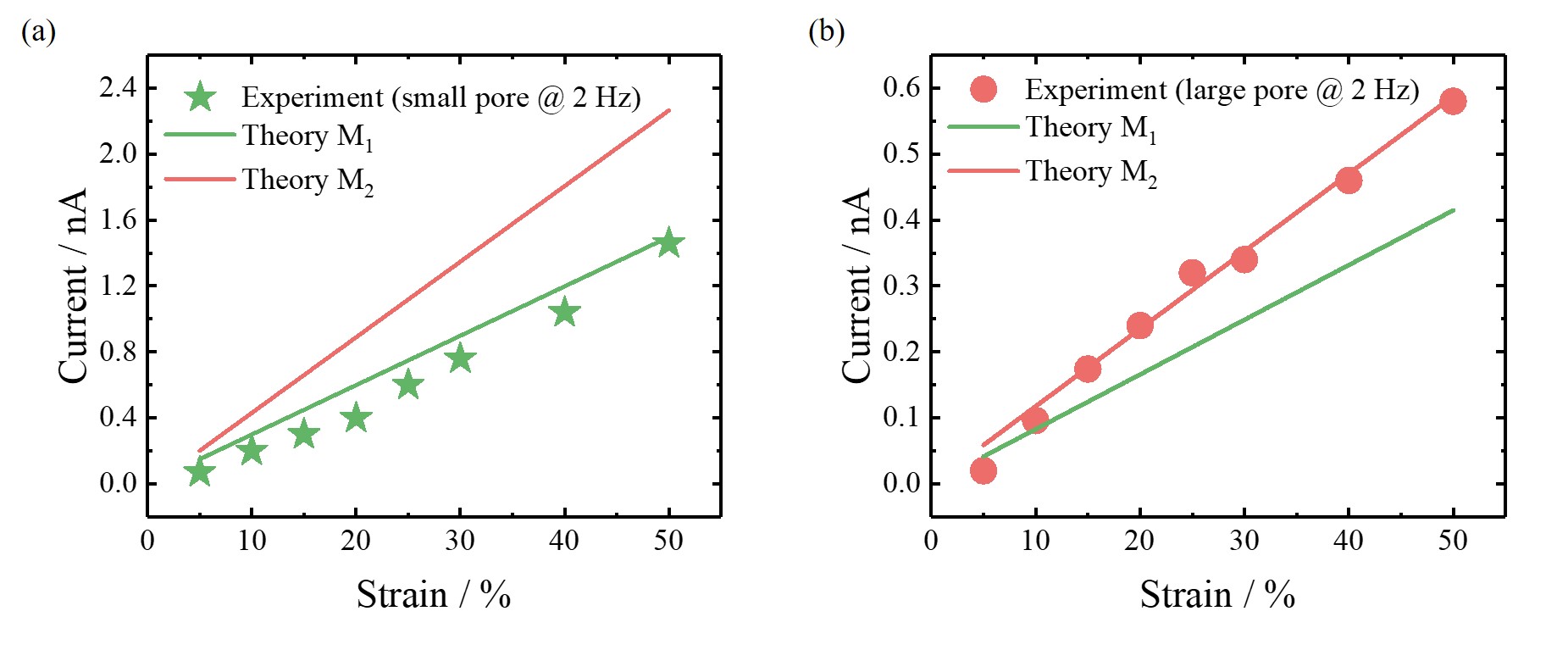}\\
  \caption{Comparison  between the experimental and theoretical  flexoelectric responses under different strains : PDMS of small pores (a) and large pores (b) under different strains at 2 Hz.}\label{Fig:10}
\end{figure}

In order to compare the efficiency of the flexoelectric response  of porous PDMS with the solid PDMS, two kinds of solid truncated pyramid PDMS specimens with the same top and bottom square surfaces ($a_1$=5 mm, $a_2$=10 mm) and different heights ( $\mathrm{H}$=6 mm and 8 mm) are fabricated as shown in the inset in Fig. 8(a), where a liner relationship can be observed between the flexoelectric-induced current and the strain gradient of the pyramids specimens. The method in Ref. \cite{32} is used to calculate the longitudinal flexoelectric coefficient $\mu_{1111}$. The measured $\mu_{1111}$ in our experiments is ca. $3.631\times10^{-10}$ C/m (the coordinate is the same as that illustrated in Fig. 2(d)), which agrees with the transverse flexoelectric coefficient $\mu_{1133}=5.3\times10^{-10}$ C/m in Ref. \cite{26}. This $\mu_{1133}$ value is used in this work.
  We define a weight and deformability specific effective coefficient ($W_\mathrm{sec}$) as $W_\mathrm{sec} ={I_\mathrm{max}}/({m_\mathrm{g} \varepsilon})$, where $I_\mathrm{max}$, $m_\mathrm{g}$ and $\varepsilon$ represent the maximum induced current, the weight of specimen and the applied macroscopic compression strain, respectively. Fig. 8(b) illustrates $W_\mathrm{sec}$ of the solid and porous specimens under the applied strain 25$\%$. The porous PDMS specimens yield two orders of magnitude higher $W_\mathrm{sec}$ than that of the truncated pyramids, and $W_\mathrm{sec}$ increases when the pore size of the porous specimen decreases.

The values of $\mu_{1131}$ and $\mu_{1133}$ are equal to each other, which can be concluded from the literature such as Refs. \cite{10,11,24}. The theoretical flexoelectric responses predicted using models $\mathrm{M}_1$ and $\mathrm{M}_2$ in section 2 and the experimental results under different frequencies and strains are shown in Fig.~9 and Fig.~10. Both theoretical models can effectively predict the flexoelectric response of the porous PDMS specimens. The theoretical results predicted by model $\mathrm{M}_1$ agree with the experimental results of the specimens of small pores, whereas the results predicted by model $\mathrm{M}_2$ agree with the experimental results of the specimens of large pore better. This may be attributed to the much larger pore diameter of the latter compared with the former.  With the large pores, the internal vertical ligaments are prone to buckling deformation (corresponding to deformation considered in $\mathrm{M}_2$). From a buckling analysis, when the aspect ratio ($l/\delta$) of beam b$_1$ is less than 3.5, it will not buckle under the maximum load in our experiment. For the porous PDMS specimens of small pores, the aspect ratio is about $3.6\pm0.6$, so M$_1$ is suitable. However, for the porous PDMS specimens of large pores, the aspect ratio is $5.3\pm0.9$, which is larger than the critical value 3.5, so M$_2$ is suitable.

\subsection{Flexoelectric response of porous PVDF}
Porous polyvinylidene fluoride (PVDF) cuboids ($5\times9\times14 \;\mathrm{mm}^3$) and beams ($40\times10\times4 \;\mathrm{mm}^3$) have been fabricated using the method  similar to that in Ref.~\cite{45}. The detailed fabrication process is given in Supplementary Material. Fig. 11(a) and (b) illustrate the picture and SEM images of the porous PVDF specimens, from which we have measured an average pore diameter of 200 $\mu$m and a ligament thickness of 60 $\mu$m. Similarly, a dynamic compression displacement load  as a sine function was applied on the porous PVDF cuboid and a dynamic bending load was applied on the porous PVDF beam. The transverse flexoelectric coefficient $\mu_{1133}=3.7\times10^{-8}$ C/m of the solid PVDF was measured by the three-point bending method, which agree with the coefficient in Refs.~\cite{24,41}. This $\mu_{1133}$ value and $\mu_{1123}=9.2\times10^{-9}$ C/m from Ref.~\cite{10} are used in the theoretical models. The experimental (dots) and theoretical (solid line) results of the flexoelectric output current of the porous PVDF cuboid at different applied macroscopic compression strains and load frequencies are shown in Fig. 11(c) and (d). Fig. 12 illustrates the experimental (dots) and theoretical (solid line) results of the flexoelectric output current of the porous PVDF beam at different applied macroscopic bending angles and load frequencies. From those figures, one can see that the flexoelectric current increases with the increase of the applied macroscopic strain, bending angle and load frequency, and the measured results agree well with the theoretical results predicted by the theoretical models.

\begin{figure}[h]
  \centering
  \includegraphics[width=12cm]{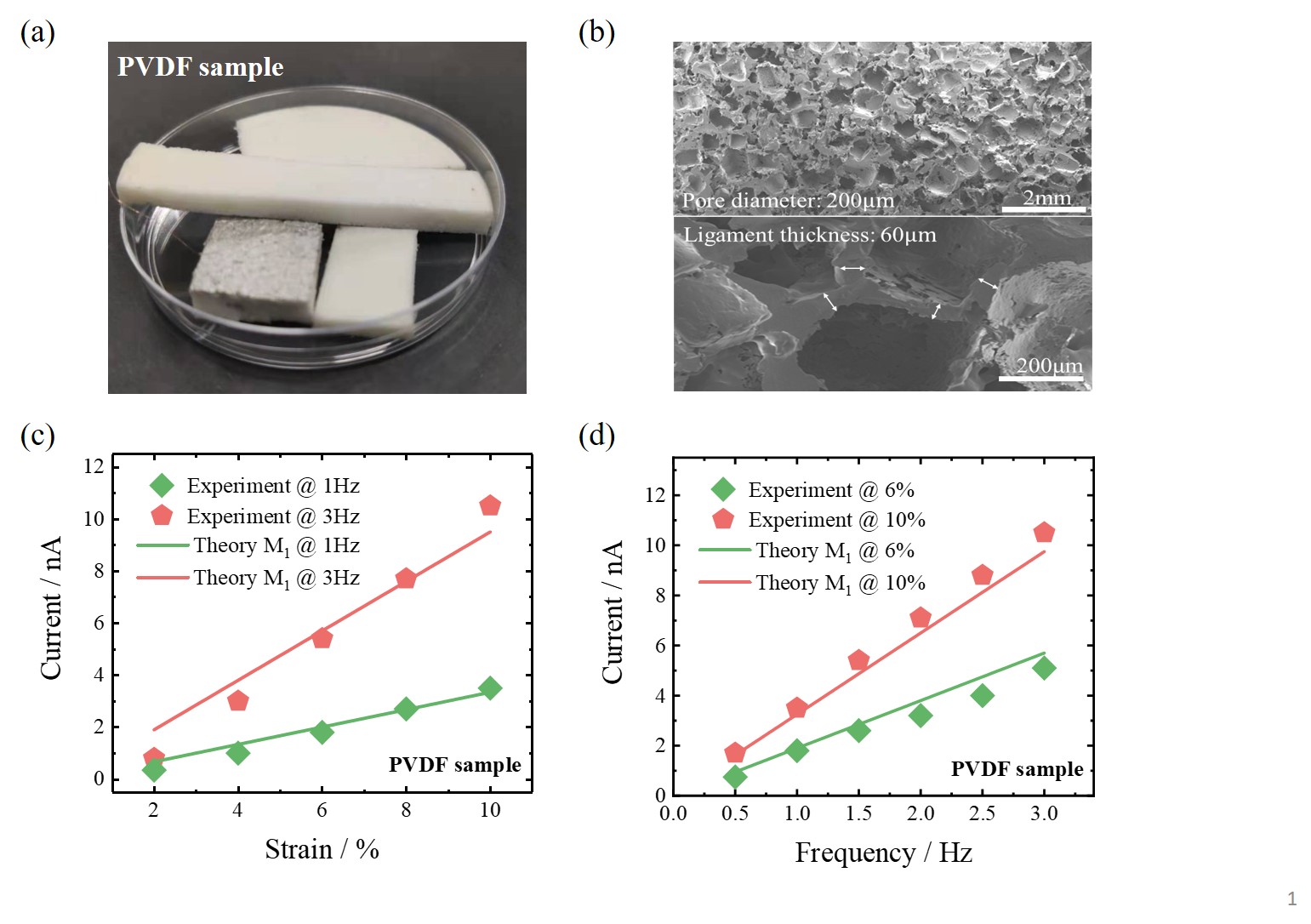}\\
  \caption{Verification of the theoretical model $\mathrm{M}_1$ with porous PVDF cuboid. (a) The picture and (b) SEM images of porous PVDF specimens. Comparison of the experimental and theoretical results under different macroscopic compressive strains (c) and  frequencies (d).}\label{Fig:11}
\end{figure}

\begin{figure}[h]
  \centering
  \includegraphics[width=7cm]{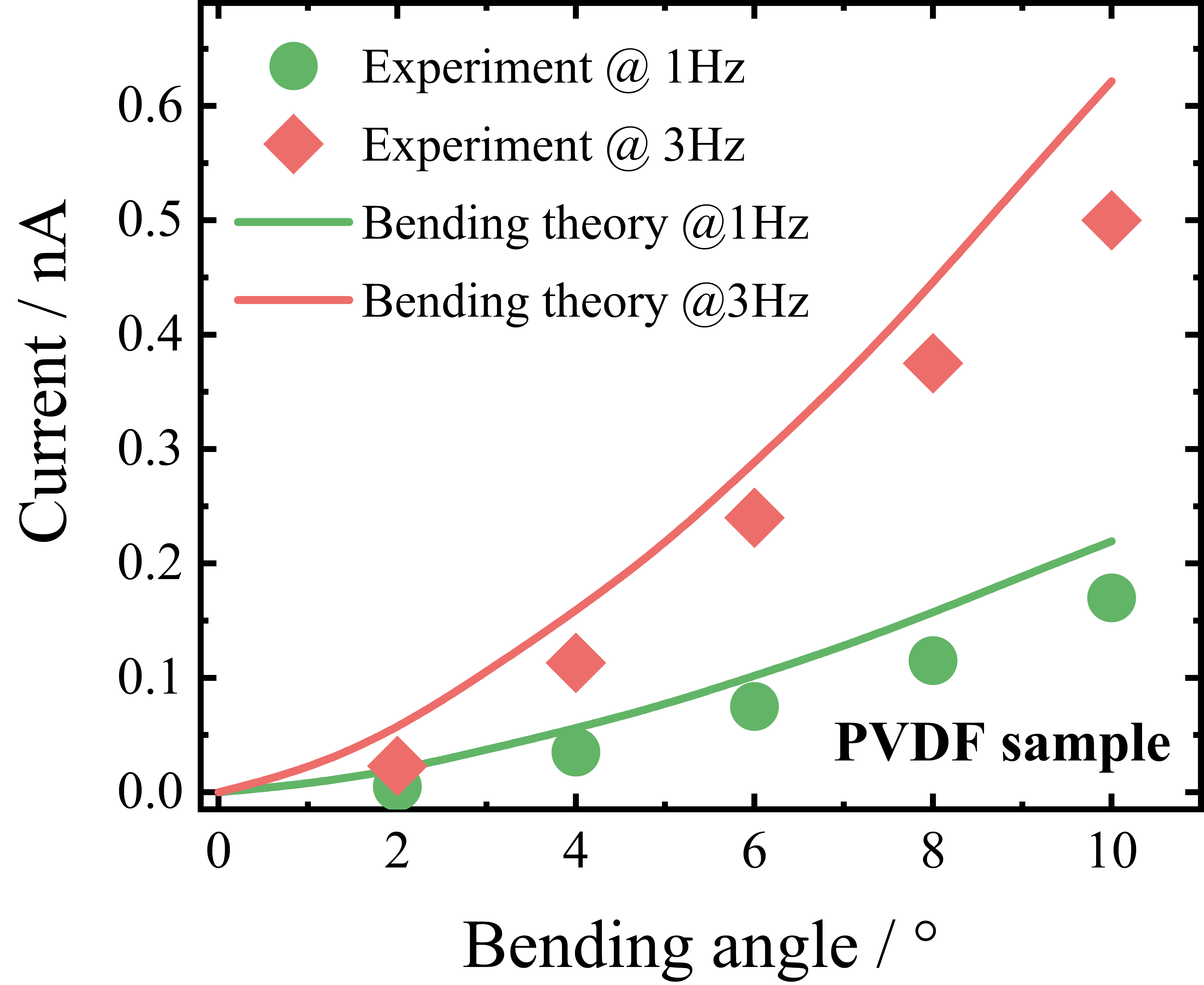}\\
  \caption{Verification of the  theoretical bending model with porous PVDF beam under different bending angles and frequencies. }\label{Fig:12}
\end{figure}

\begin{figure}[h]
  \centering
  \includegraphics[width=14cm]{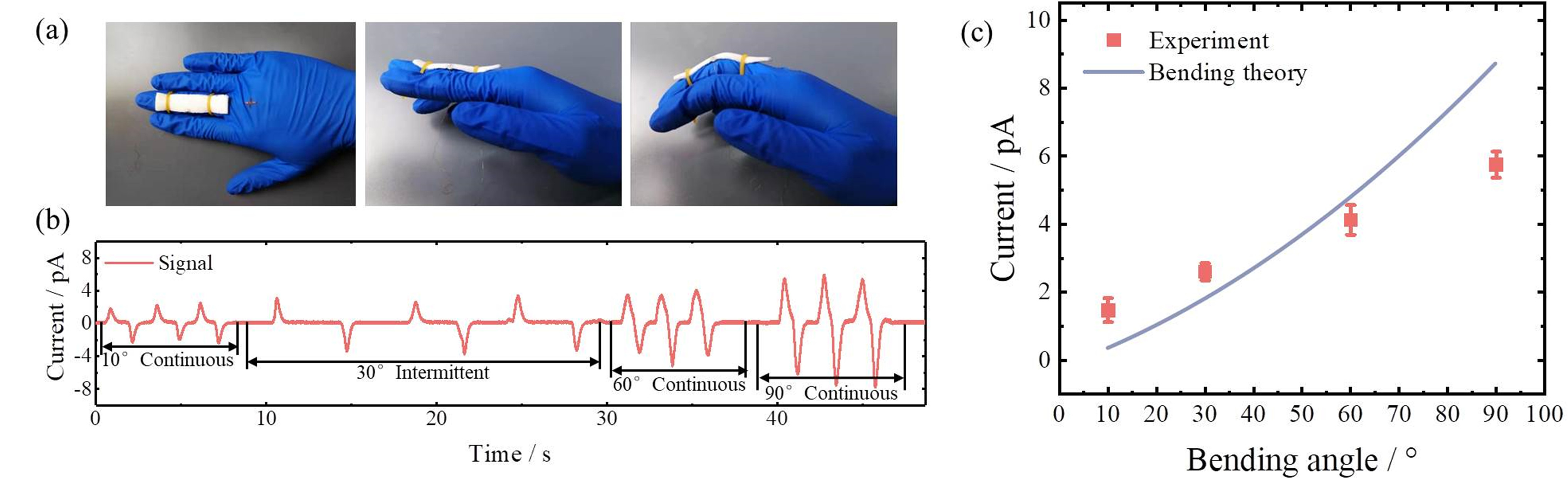}\\
  \caption{A bending sensor of the thin porous PDMS film ($60\times15\times5 \;\mathrm{mm}^3$) sensing the knuckle bending movement (a), the flexoelectric output at different bending angles and modes (b), and the relationship between the flexoelectric-induced current and bending angles (c).} 
\end{figure}

\subsection{Flexible bending sensor}

Benefiting from the high flexoelectric response, diverse macrostructures, unlimited loading forms, high flexibility and breathability, porous polymers may enable their applications as wearable sensors. Here, as a final verification, and also practical application, of our model, we design a bending sensor
made of a thin porous PDMS film ($60\times15\times5 \;\mathrm{mm}^3$), as shown in Fig. 13, which senses the knuckle bending movement. The porous PDMS with point-like electrodes $(6\times9 \;\mathrm{mm}^2)$ on both sides of the midpoint of the film is tied on top of the middle finger by two rubber bands, and the ends of the film do not cross the first and the end finger joint as shown in Fig. 13(a). Fig. 13(b) shows the flexoelectric responses of the film at different bending angles and modes, where ``$10^{\circ}$ continuous'', ``$60^{\circ}$ continuous'', ``$90^{\circ}$ continuous'' and ``30$^{\circ}$ intermittent'' represent the repeated bending and releasing with bending angles of 10$^{\circ}$, 60$^{\circ}$, 90$^{\circ}$ and the intermittently bending and releasing with the bending angle of 30$^{\circ}$, respectively. It is seen that the porous PDMS sensor can measure the knuckle bending angle and action mode accurately and effectively. The excellent liner relationship between the flexoelectric-induced current and bending angles are illustrated in Fig. 13(c), which demonstrates the promising applications of porous PDMS in bending sensors based on flexoelectricity. The theoretical flexoelectric current based on the bending model established in section 2 is also shown in Fig. 13(c). The theoretical results are consistent with the experimental measurements, which further verifies the rationality of our theoretical model in this more practical test.

 The two bulk polymers PDMS and PVDF ($\alpha$-PVDF) we used in this work do not have piezoelectric effect, though $\beta$-PVDF is a kind of widely used piezoelectric material~\cite{martins2014}; however, their porous specimens exhibit the piezoelectric-like effect under macroscopic strains, due to flexoelectricity at the microscopic scale. Thus, we can define and calculate the {\it equivalent} piezoelectric strain coefficients of the porous PDMS and $\alpha$-PVDF specimens under compression: 5.16~pC/N and 3.5~pC/N, compared to those ($d_{33}$) of lead zirconium titanate PZT (ceramic) and $\beta$-PVDF (polymer): 410~pC/N~\cite{saitoetal2004} and  and 24-34 pC/N~\cite{martins2014}. The former are much smaller than the latter. However, when we use weight specific values of the piezoelectric strain coefficients, that is, the piezoelectric strain coefficients divided by the densities. The {\it equivalent} weight specific piezoelectric strain coefficients of the porous PDMS and $\alpha$-PVDF specimens are 14.41 pC$\cdot$ml$\cdot$N$^{-1}$$\cdot$g$^{-1}$ and 4.01~pC$\cdot$ml$\cdot$N$^{-1}$$\cdot$g$^{-1}$, whereas those of lead zirconium titanate PZT (ceramic) and $\beta$-PVDF (polymer) are 51.2 pC$\cdot$ml$\cdot$N$^{-1}$$\cdot$g$^{-1}$ and 11.4-16.2~pC$\cdot$ml$\cdot$N$^{-1}$$\cdot$g$^{-1}$, respectively. Therefore, the {\it equivalent} weight specific piezoelectric strain coefficients of the porous PDMS and $\alpha$-PVDF specimens are comparable with those of the widely used PZT and $\beta$-PVDF.

\section{Conclusions}
We establish the flexoelectric model of interconnected porous structures. The theoretical results show that the flexoelectric response increases with the decrease of the pore size and the increase of the ligament thickness, and also increases with the decrease of the Poisson ratio. Then, we verify the theoretical model with the porous PDMS and porous PVDF specimens by measuring their flexoelectric response under compression and bending. The  weight and deformability specific flexoelectric response ($W_\mathrm{sec}$) of the porous PDMS is two orders of magnitude higher than that of the solid truncated pyramid PDMS. Finally, we demonstrate the possible applications of both the theory and the porous media in light-weight sensitive and flexible sensors.
In this work, only up to 50$\%$ compressive strain are applied on the porous structures, larger  strains and other loading forms such as tension and torsion can also be implemented. From a broader perspective, this study highlights the potential of porous materials in the field of flexoelectric devices (sensing, actuating, energy harvesting and so on) and possible rational micron/nano structural design with various materials to achieve expected flexoelectric responses.

\begin{flushleft}
\textbf{\emph{Acknowledgements}}
\end{flushleft}

The authors thank Professor Shengping Shen and Associate Professor Qian Deng of Xi'an Jiao Tong University,  and Professor Haixia Zhang, Dr. Qiancheng Zhao, Dr. Jian Cui, and
 Dr. Jingeng Mai of Peking University for helpful discussions. We also thank Professor Anyuan Cao of Peking University for help with the experiment. We are grateful to the anonymous reviewers whose incisive comments have helped to improve the technical quality of the work. This work is supported by the National Natural Science Foundation of China (NSFC No. 11572051, 11890681 and 11521202).

\begin{flushleft}
\textbf{\emph{Author contributions}}
\end{flushleft}

L.H.S. and J.W. had the original ideas. L.H.S. conceived and designed the project. M.Z. and D.Y. carried out the whole experiment. M.Z. and D.Y. drafted the manuscript and L.H.S. and J.W. contributed to the final version of the manuscript. All authors discussed the results and commented on the manuscript.

\begin{flushleft}
\textbf{\emph{Competing interests}}
\end{flushleft}

The authors declare no competing interests.

\begin{flushleft}
\textbf{\emph{Data and materials availability}}
\end{flushleft}

The data that support the findings of this study are available from the corresponding author upon reasonable request.

\end{document}